\documentclass{article}

\usepackage{spconf,amsmath,epsfig}
\usepackage{geometry}                
\usepackage{graphicx}
\usepackage{amssymb}
\usepackage{epstopdf}
\usepackage{tikz}
\usepackage{pgfplots}
\usepackage{url}
\usepackage{scalefnt}
\usepackage{setspace}
\title{Data management and analysis with WRF and SFIRE}
\name{Jonathan D. Beezley$^{\star}$, Mavin Martin$^{\dagger}$, Paul Rosen$^{\dagger}$, Jan Mandel$^{\star}$, Adam K. Kochanski$^{\ddagger}$\thanks{
Supported by NSF grant AGS-0835579 and NIST Fire Research Grants Program grant 60NANB7D6144.}}
\address{$^{\star}$Department of Mathematical and Statistical Sciences, University of CO Denver\\
$^{\dagger}$Scientific Computing and Imaging Institute, University of Utah\\
$^{\ddagger}$Department of Meteorology, University of Utah}

\begin{document}
\maketitle

\begin{abstract}
We introduce several useful utilities in development for the creation and analysis of real wildland fire simulations
using WRF and SFIRE.  These utilities exist as standalone programs and scripts as well as extensions to 
other well known software.  Python web scrapers automate the process of downloading and preprocessing 
atmospheric and surface data from common sources.  Other scripts simplify the domain setup by creating
parameter files automatically.  Integration with Google Earth allows users to explore the simulation 
in a 3D environment along with real surface imagery.  Postprocessing scripts provide the user with a number
of output data formats compatible with many commonly used visualization suites allowing for the creation
of high quality 3D renderings.  As a whole, these improvements build toward a unified web application
that brings a sophisticated wildland fire modeling environment to scientists and users alike.
\end{abstract}


\section{Introduction}

The weather research and forecasting model (WRF) is a full-featured mesoscale weather model used as a research
tool and operational forecasting \cite{Michalakes-2004-WRF}.  SFIRE is an add-on wildland fire forecasting model
coupled with WRF \cite{Mandel-2011-CAF}. An earlier version is available in WRF release as 
WRF-Fire.\footnote{\url{http://www.openwfm.org/wiki/WRF-Fire_development_notes}}
Because WRF is widely used in the research community, there are a large
number of standard tools available for creating and visualizing its simulation outputs.  However, SFIRE acts on 
domains that are significantly higher resolution than the typical WRF simulation.  In addition, SFIRE uses 
surface variables on a refined subgrid that is different from the standard atmospheric surface variables.  
These subgrid variables and the higher resolution modeling typically requires special processing for input 
and visualization that is not supported by most commonly used tools.  For this reason, we have begun to develop
new software that mimic and extend standard utilities for data processing and visualization specifically tailored
for use with SFIRE.

\section{Model initialization and data preprocessing}

As computers become more powerful, there is increasing interest in simulating smaller scale phenomena
using WRF than is
typically associated with mesoscale weather forecasting.  
In particular, a typical SFIRE simulation occurs 
at mesh resolutions on the order of 10 m or less.  Even the highest resolution surface data provided with WPS 
is several hundred times coarser.  For these fine scale domains, the ability to import custom datasets into
WPS is essential for the initialization of a realistic simulation \cite{Jordanov-2011-SWF}.
Resources such as the USGS's seamless data server that provide open access to high quality 
surface data are generating a large interest in initializing WRF simulations using custom datasets.
WPS provides a mechanism for these datasets through a simple binary file format that is described
in the WRF technical documentation \cite{Wang-2010-AUG}; however, there is no standard API or GIS software capable of 
writing to it. Users who lack sufficient technical knowledge are currently unable to process this
data into WRF.  

Prior work related to WRF-Fire has lead to small utilities that are able to convert standard GIS GeoTIFF 
files into Geogrid's binary file format.  
The GeoTIFF specification is commonly used 
and can be provided as output from a wide range of GIS applications. 
TopoGrabber\footnote{\url{http://laps.noaa.gov/topograbber}} 
is a Python application based on this work
that is capable of downloading and converting topological data automatically. 
More recently, modifications to Geogrid have been written allowing it to read GeoTIFF files
directly, and extensions to WPS standard programs have made it possible to 
ingest GeoTIFF data directly from the USGS into a WRF simulation \cite{Beezley-2011-IHS}.
The implemented changes to Geogrid allows a user to import a USGS GeoTIFF dataset directly into the WRF workflow without the
(often) difficult and error prone procedure of converting the data.  
The GeoTIFF interface is compiled in as an optional component 
and is useful for both experienced GIS users with complicated workflows and those who just want to replace a single data source.  
The implementation allows overriding erroneous metadata without modification to the source images, and efficient access to the
data.  

With tools in place to automatically retrieve and integrate known data sources into a simulation, it is now possible 
to automate nearly all of the steps necessary to initialize a simulation for WRF and SFIRE.  A command-line client
has been written using Python to generate all required parameter and input files for a simulation given user
defined domain and fire ignition specifications.  This script is currently being integrated with a web application
based on Google Maps, providing users with a fully graphical environment for designing a simulation on real surface
imagery.  This client application is coupled with a server located on a high performance compute cluster that 
runs the simulation on demand and returns results back to the client as they are processed.

\section{Model evaluation and data postprocessing}

By default, WRF generates NetCDF binary files containing the raw simulation output.  These files
contain a large number of variables that are useful in analyzing the simulation output; however,
they are often too large to be transferred to the user's computer for visualization.  Several utilities
have been written that allow users to visualize the simulation without downloading the full output.
For a quick
overview of the fire propagation, it is possible to generate a KMZ file containing a raster image
of any surface variable or a georeferenced polygon outlining the burning region at a given timestep.
This KMZ file can be opened in a number of GIS applications including Google Earth to be overlaid
on a rendering of the earth's surface.  Google Earth allows the user to play back the images in
the KMZ files as an animation with a timestamp indicating the simulation time.  The web application
in development uses this capability to display the results of the running simulation on top of 
Google Maps.

There are a number of options for rendering a 3D visualization of the simulation.  VAPOR is an application
developed at NCAR designed for visualization of standard WRF atmospheric simulations.  The postprocessing
utility \texttt{wrf2vdf} takes a standard WRF NetCDF output file and generates a more compact vdf file
containing only the fields the user wishes to investigate.  VAPOR features a very fast and easy to use GUI
for generation of high quality images; however, it does not support the refined grids used by the fire
variables.  When using VAPOR with SFIRE, users are limited to low resolution versions of fire variables,
which can result in blocky images.  

Other popular 3D visualization applications include Paraview, MayaVI, and VisTrails.  These applications
all use the Visualization Toolkit (VTK) as a backend and do not have any built in capability for reading
WRF output files.  SFIRE contains a Python script \texttt{wrf2vtk} that can generate VTK compatible 
files from WRF outputs at full resolution.  This script is much like VAPOR's \texttt{wrf2vdf} in that
it converts the data to a more compact representation and compresses the files for transfer to a
local computer. 

\section{Toward a fire forecasting web portal}

Automated scripting of the model setup procedure has lead to the development of a web portal providing
users with a simple interface for running SFIRE simulations using real data.  The user registers with
the site with a username and password.  Upon logging in, he or she is presented with a Google Maps
interface with instructions to click on a point on the map to create a new ignition.  When clicked,
a dialog box appears allowing the user to customize the parameters of the simulation.  The user is 
notified by email on completion of the simulation when the user can return to the web portal to view
the output of this and all prior simulations.  Currently, the output of the simulation displays as an
animation the heat flux into the atmosphere from the fire.  In the future, we plan to add the ability
to visualize other surface variables such as fuel types, winds, and fire danger ratings.

The web frontend is built
on top of a Google Maps interface using its javascript interface.  The core web application including
user and content management is served by Django.  The backend component is executed at a remote site on a 
compute cluster accessible via ssh from the web server.  The web application passes a string in the form of
simple keyword=value pairs specifying parameters of the fire simulation.  These parameters include the
ignition location and time chosen by the user, domain size and resolution, and simulation run time among
others.  The web server executes an application script on the compute server providing the simulation
parameters and a session ID.  The compute server initializes and executes the simulation while a post
processing daemon watches for new output files.  Whenever a new output file is detected, a visualization 
script is executed generating a KMZ file, which is then uploaded to the web server to be displayed to the
user.

\begin{figure*}[htb]
\begin{minipage}[b]{1.0\linewidth}
\centering
\centerline{\includegraphics[width=6in]{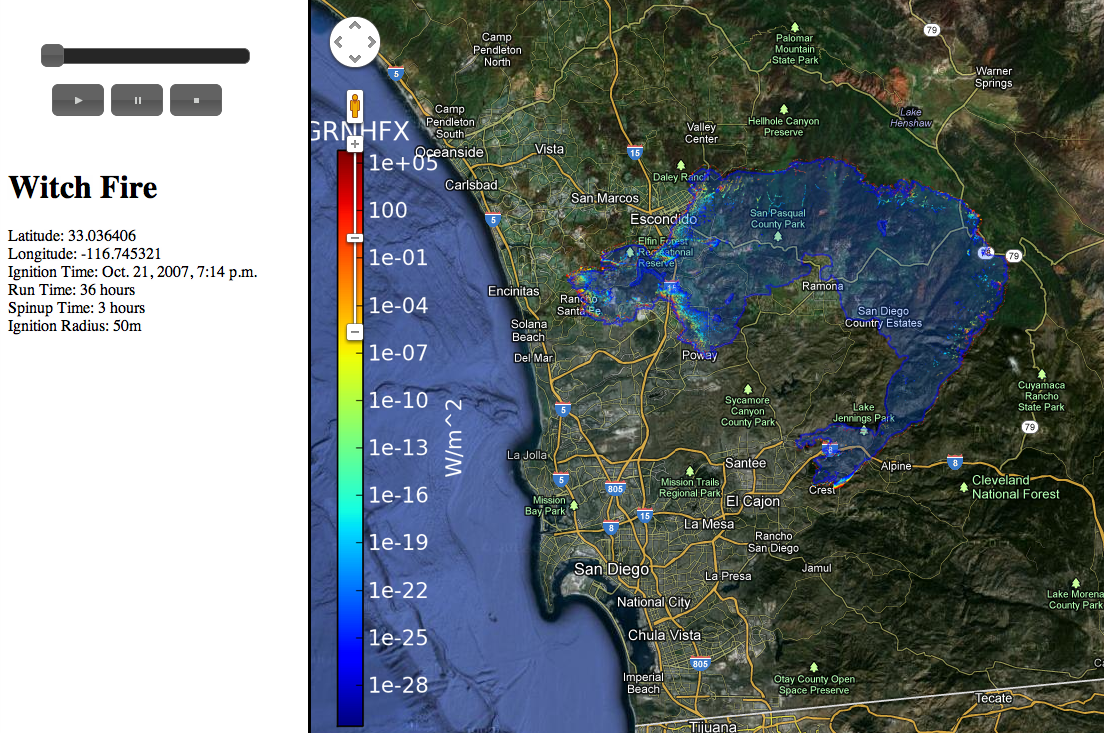}}
\caption{SFIRE web portal results page.  Along with a standard Google Maps interface, the results portal 
contains animation controls so the user can view the development of the fire front.  Prior simulations
are stored on the user's account page and can be shared with other users.}
\medskip
\end{minipage}
\end{figure*}

\section{Conclusion and future work}

The utilities presented here are designed to enhance the usability of SFIRE for the average user 
by eliminating the need for complex data conversion or an in depth understanding of the underlying
code.  They
are leading to a unified application where a user can design a simulation through a series of 
graphical dialogs, spawn the simulation on a remote compute cluster, and visualize the output of the
running simulation locally.  The development of such an application would provide an invaluable
tool for research and education.

\bibliographystyle{IEEEbib}
\bibliography{igarss12}

\end{document}